\documentclass[aps,prd,showpacs,graphicx,twocolumn,,superscriptaddress,10pt]{revtex4-1}%
\usepackage{mathrsfs}
\usepackage{times}
\usepackage{amsmath}
\usepackage{graphicx}
\usepackage{sidecap}
\usepackage{txfonts}
\usepackage{color}
\usepackage{ulem}
\usepackage[colorlinks=true,citecolor=blue,linkcolor=magenta]{hyperref}
\usepackage{subeqnarray}
\usepackage{verbatim}
\usepackage{epstopdf}
\usepackage{cases}
\usepackage[symbol]{footmisc}
\usepackage{subfigure}
\usepackage{epsfig}

\DefineFNsymbols{nu}{\dag *}\setfnsymbol{nu}
\renewcommand{\aa}{\mathring{a}}

\begin{document}

\title{Consistent Lorentz violation features from
	near-TeV IceCube neutrinos}

\author{Yanqi Huang}
\affiliation{School of Physics and State Key Laboratory of Nuclear Physics and
	Technology, Peking University, Beijing 100871,
	China}

\author{Hao Li}
\affiliation{School of Physics and State Key Laboratory of Nuclear Physics and
	Technology, Peking University, Beijing 100871,
	China}

\author{Bo-Qiang Ma}
\email{Corresponding author: mabq@pku.edu.cn}
\affiliation{School of Physics and State Key Laboratory of Nuclear Physics and
	Technology, Peking University, Beijing 100871,
	China}
\affiliation{Collaborative Innovation Center of Quantum Matter, Beijing, China}
\affiliation{Center for High Energy Physics, Peking University, Beijing 100871, China}

\date{\today }

\begin{abstract}
A recent proposal to associate 60~TeV to 2~PeV IceCube neutrino events with gamma-ray bursts~(GRBs) indicates the Lorentz violation of cosmic neutrinos and leads further to the $CPT$ symmetry violation between neutrinos and antineutrinos. Here we find that another 12 northern hemisphere track events possibly correlated with GRBs from three-year IceCube data satisfy the same regularity at a lower energy scale around 1~TeV. The combined fitting indicates a Lorentz violation scale ${E}_{\rm LV}=(6.4\pm 1.5)\times10^{17}~{ \rm GeV}$ and an intrinsic time difference ${\Delta {t}_{\rm in}=(-2.8\pm 0.7)\times10^2~{\rm s}}$,
from which we find an earlier emission of neutrinos than photons at the GRB source.
We also suggest analyzing neutrino events detected a few minutes before the GRB trigger time to test the $CPT$ violation of ultrahigh-energy neutrinos.
\end{abstract}

\maketitle


\section{Introduction}
Cosmic neutrinos from astrophysical sources have the potential to reveal novel features of the Universe. The IceCube Neutrino Observatory has observed plenty of neutrino events, including dozens of neutrinos with energies above 30~TeV, with also four PeV scale neutrinos~\cite{Aartsen:2013bka,Aartsen:2013jdh,Aartsen:2014gkd,Kopper:2015vzf,Aartsen:2016ngq,Aartsen:2018vtx}. It has long been expected that gamma-ray bursts~(GRBs) can be one class of sources for ultrahigh-energy cosmic neutrinos~\cite{Eichler:1989ve,Paczynski:1994uv,Waxman:1995vg,Waxman:1997ti,Piran:1999bk}. However, the IceCube Collaboration examined GRBs within a short
temporal window of a few hundred seconds and
suggested possible correlations between GRBs and some lower-energy neutrinos with energies around 1~TeV (we call ``near-TeV'' later), with the conclusion that these near-TeV events are also consistent with atmospheric backgrounds~\cite{Aartsen:2014aqy,Aartsen:2016qcr,Aartsen:2017wea}.

The situation changes significantly when combining the association between neutrinos and GRBs with studies on Lorentz-invariance violation~(LV)~\cite{AmelinoCamelia:1997gz,Jacob:2006gn}.
The long propagation distance and the high energy of GRB neutrinos can produce an observable difference between GRB trigger time and the arrival time of neutrinos due to the LV effect~\cite{Jacob:2008bw}. Amelino-Camelia and collaborators associated IceCube 60-500~TeV shower neutrino events with GRB candidates within the time range of three days, and revealed roughly compatible speed variation features between GRB neutrinos~\cite{Amelino-Camelia:2015nqa,Amelino-Camelia:2016fuh,Amelino-Camelia:2016ohi} and photons~\cite{Shao:2009bv,Zhang:2014wpb,Xu:2016zxi,Xu:2016zsa,Xu:2018ien,Liu:2018qrg}. In a recent study~\cite{Huang:2018}, all four PeV scale IceCube neutrino events are associated with GRBs by extending the temporal window to a longer range of three months, and such four events are found to be consistent with TeV scale events for an energy-dependent speed variation. Such findings indicate the Lorentz violation of cosmic neutrinos.
It is also found that there are both time ``delay'' and ``advance'' events, which can be explained by different propagation properties between neutrinos and antineutrinos~\cite{Huang:2018}. This leads further to the charge conjugation, parity transformation, and time reversal ($CPT$) symmetry violation between neutrinos and antineutrinos, or an asymmetry between matter and antimatter~\cite{Zhang:2018otj}. As these results are of fundamental importance, it is thus necessary to check whether additional IceCube neutrino events still support the revealed regularity~\cite{Amelino-Camelia:2016ohi,Huang:2018} or not.

In this work, we provide a reexamination of these near-TeV neutrino events which are possibly associated with GRBs by the IceCube Collaboration~\cite{Aartsen:2016qcr,Aartsen:2017wea}.
We find that all of these near-TeV northern hemisphere ``track'' events with associated GRBs during 2012--2015 can satisfy the regularity obtained from the 60~TeV to 2~PeV neutrinos with associated GRBs~\cite{Amelino-Camelia:2016ohi,Huang:2018}. By performing the combined linear fitting of the 13 high-energy TeV and PeV events~\cite{Amelino-Camelia:2016ohi,Huang:2018} (with energies from 60~TeV to 2~PeV) together with the 12 lower-energy near-TeV northern hemisphere events~\cite{Aartsen:2014aqy,Aartsen:2016qcr,Aartsen:2017wea} (with energies around 1~TeV), we find that these lower-energy GRB neutrinos are consistent with the TeV and PeV GRB neutrinos for an energy-dependent speed variation at $E_{\rm LV}=(6.4\pm 1.5)\times10^{17}~{\rm GeV}$.
We also find that GRB neutrinos are emitted $280\pm 70$~seconds before GRB photons at the source. As a prediction, we suggest analyzing neutrino events observed several minutes before the GRB trigger time to test the revealed regularity.

\section{Model}
The LV physics can be determined or limited by the energy-dependent speed variation of GRB photons and neutrinos~\cite{AmelinoCamelia:1997gz,Jacob:2006gn}. For a particle propagating in the quantum spacetime with energy $E\ll E_{\rm Pl}$~(the Planck scale $E_{\rm Pl}\approx1.22\times10^{19}~{\rm GeV}$), the leading terms in a Taylor series expansion of the LV modified dispersion relation can be written as
\begin{equation}
v(E)=c\left[1-s_n\frac{n+1}{2}\left(\frac{E}{E_{{\rm LV},n}}\right)^n\right],
\end{equation}
where $n=1$ or $n=2$ corresponds to linear or quadratic dependence of the energy, $s_n=\pm1$ is the sign factor of LV correction, and $E_{{\rm LV},n}$ is the $n$th-order LV scale to be determined by experiments.
Such a speed variation can cause a propagation time difference between particles with different energies. By taking into account the cosmological expansion, the LV time correction in the $n=1$ case of two particles with energies $E_{\rm h}$ and $E_{\rm l}$ , respectively, can be written as~\cite{Jacob:2008bw,Ellis:2002in}
\begin{equation}
\Delta t_{\rm LV}=s\cdot(1+z)\frac{K}{E_{\rm LV}},
\end{equation}
where $z$ is the redshift of the GRB source, $s=\pm1$ is the sign factor and
\begin{equation}
K=\frac{E_{\rm h}-E_{\rm l}}{H_0 }\frac{1}{1+z}\int^z_0\frac{(1+z^\prime) {\rm d} z^\prime}{\sqrt{\Omega_{\rm m}(1+z^\prime)^3+\Omega_\Lambda}},
\label{LV factor}
\end{equation}
is the LV factor. We adopt the cosmological parameters~\cite{Agashe:2014kda} $[\Omega_{\rm m},\Omega_\Lambda]=[0.315^{+0.016}_{-0.017},0.685^{+0.017}_{-0.016}]$ and the Hubble expansion rate $H_0=67.3\pm 1.2~{\rm km\cdot s^{-1}\cdot Mpc^{-1}}$.  By taking the intrinsic time difference $\Delta t_{\rm in} $ into consideration, the observed arrival time difference $\Delta t_{\rm obs}$ between two particles detected on the Earth is
\begin{equation}
\frac{\Delta t_{\rm obs}}{1+z}=\Delta  t_{\rm in}+s\cdot\frac{K}{E_{\rm LV}}.
\label{linear relation}
\end{equation}
For neutrinos emitted with associated GRBs, $\Delta t_{\rm obs}$ can be represented by the difference between the arrival time of the neutrino and the trigger time of the given GRB.
According to Eq.~(\ref{linear relation}), there would be a linear relation between $\Delta t_{\rm obs}/(1+z)$ and $K$, if the energy-dependent speed variation does exist.

\begin{table}[t]
	\small
	\centering
	\caption{\small\sf Parameters of TeV and PeV GRB neutrinos. The 13 GRB candidates are suggested from the associated GRBs of IceCube neutrinos with energies above 60~TeV by the maximum correlation criterion~\cite{Amelino-Camelia:2016fuh,Amelino-Camelia:2016ohi,Huang:2018}.
		12 of the 13 events are ``shower" events and the last 2.6~PeV one is ``a track" event. The event serial numbers here are provided by the IceCube database, except the last ATel id  $\#7856$.
		The mark $^*$ represents the estimated value of the redshift.}
	\label{tab:1}
    \begin{ruledtabular}
	\centering
	\begin{tabular}{lllll}
		\noalign{\vspace{0.5ex}}
		Event & GRB & $z$ & $\Delta t_{\rm obs}~(10^3~\rm s)$ & $E$~(TeV)   \\
		\hline
		\noalign{\vspace{0.5ex}}
		$\#2$& 100605A&$1.497^*$&$-$113.051&117.0   \\
		
		\noalign{\vspace{0.5ex}}
		$\#9$&  110503A&1.613&80.335&63.2 \\
		
		\noalign{\vspace{0.5ex}}
		$\#11$& 110531A&$1.497^*$&185.146&88.4  \\
		
		\noalign{\vspace{0.5ex}}
		$\#12$& 110625B&$1.497^*$&160.909&104.1 \\
		
		\noalign{\vspace{0.5ex}}
		$\#14$& 110725A&$2.15^*$&1320.217&1040.7   \\
		
		\noalign{\vspace{0.5ex}}
		$\#19$&  111229A&1.3805&73.960&71.5   \\
		
		\noalign{\vspace{0.5ex}}
		$\#20$& 120119C&$2.15^*$&$-$1940.176&1140.8   \\
		
		\noalign{\vspace{0.5ex}}
		$\#26$& 120219A&$1.497^*$&229.039&210.0  \\
		
		\noalign{\vspace{0.5ex}}
		$\#33$& 121023A&$0.6^*$&$-$171.072&384.7  \\
		
		\noalign{\vspace{0.5ex}}
		$\#35$& 130121A&$2.15^*$&$-$2091.621&2003.7   \\
		
		\noalign{\vspace{0.5ex}}
		$\#40$& 130730A&$1.497^*$&$-$179.641&157.3   \\
		
		\noalign{\vspace{0.5ex}}
		$\#42$& 131118A&$1.497^*$&$-$146.960&76.3   \\
		
		\noalign{\vspace{0.5ex}}
		$\#7856$ &140427A&$2.15^*$&3827.439  &  $2.6\times10^3$  \\
		
	\end{tabular}
	\end{ruledtabular}

\end{table}

\begin{figure}[t]
	\small
	\includegraphics[width=9cm]{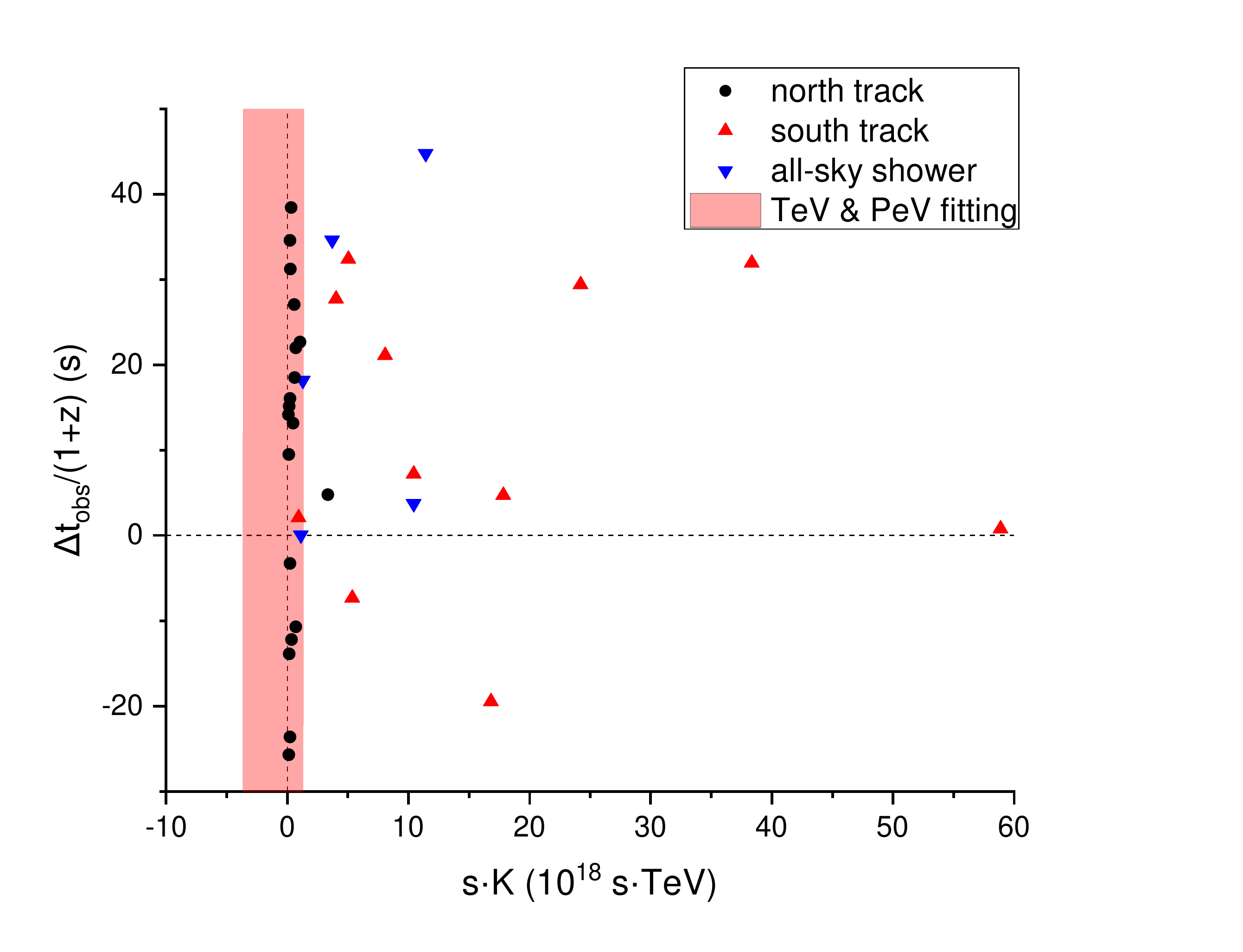}
	\caption{\small   $\Delta t_{\rm obs}/(1+z)$ versus the $K$ plot for near-TeV neutrino events. Points~(black) are northern hemisphere track events from seven-year data~\cite{Aartsen:2014aqy,Aartsen:2017wea}, up triangles~(red) are southern hemisphere track events from the five-year search~\cite{Aartsen:2017wea}, and down triangles~(blue) are shower events from the three-year search for all flavors~\cite{Aartsen:2016qcr}. The red colored region is the error band of the speed variation regularity from TeV and PeV GRB neutrinos~\cite{Amelino-Camelia:2016ohi,Huang:2018}. All of the northern hemisphere events detected during 2012--2015 fall within the error band.
	}\label{fig1}
\end{figure}

\section{TeV and PeV scale GRB neutrinos}
A regularity of energy-dependent speed variation was revealed from the studies on TeV and PeV scale IceCube neutrino events with associated GRBs~\cite{Amelino-Camelia:2016ohi,Huang:2018}. Amelino-Camelia and collaborators~\cite{Amelino-Camelia:2016fuh,Amelino-Camelia:2016ohi} selected 9 GRB candidates from the associated GRBs of IceCube TeV neutrino events with the time mismatch of three days. These 9 TeV events are selected from the shower events detected from 2010 to 2014 and with energies between 60 and 500~TeV. In a recent study~\cite{Huang:2018}, all 4 PeV neutrinos are found to be associated with GRB candidates in the time range within three months. The associations of these 13 neutrino events with GRBs are based on the maximum correlation together with direction and time criteria~\cite{Amelino-Camelia:2016ohi,Huang:2018}. These 13 events with the associated GRBs, the redshift values $z$, the observed time differences $\Delta t_{\rm obs}$, and the neutrino energies $E$ are listed in Table~\ref{tab:1}. The observed time difference $\Delta t_{\rm obs}$ could be positive or negative, just like the LV time correction $\Delta t_{\rm LV}$. Following the convention in Refs.~\cite{Amelino-Camelia:2016ohi,Huang:2018}, we call $\Delta t_{\rm obs}>0$~(or $<0$) events ``late~(or early) neutrinos,'' and call $\Delta t_{\rm LV}>0$~(or $<0$) cases ``time delay~(or advance).''

Since the redshift measurement is not available for most
GRBs, we use "most likely�estimated values of the redshift
for those GRBs without redshift measurement~\cite{Amelino-Camelia:2016ohi,Huang:2018}. We find a well linear correlation between the observed time difference and the LV factor, which can be described by a pair of inclined lines~\cite{Huang:2018}
\begin{equation}
|\frac{\Delta t_{\rm obs}}{1+z}-\Delta t'_{\rm in}|=\frac{K}{E'_{\rm LV}},
\label{abs.}
\end{equation}
with the slope and the intercept
\begin{eqnarray}
1/E'_{\rm LV}&=&(1.53\pm0.10)\times10^{-15}~{\rm TeV^{-1}},\label{slope1}\\
\Delta t'_{\rm in} &=&(1.7\pm3.6)\times10^3~{\rm s}.\label{intercept1}
\end{eqnarray}
The correlation coefficient is $r=0.978$, which implies a relatively strong linear correlation between $\Delta t_{\rm obs}$ and $K$. Such regularity indicates a Lorentz violation scale $E'_{\rm LV}=(6.5\pm0.4)\times 10^{17}~\rm{GeV}$, which is comparable with the results determined by GRB photons~\cite{Zhang:2014wpb,Xu:2016zsa,Xu:2016zxi,Amelino-Camelia:2016ohi,Xu:2018ien}. The error range of $\Delta t_{\rm in}$ covers the zero point; hence, whether these neutrino events are emitted before or after the GRB photons cannot be determined by only analyzing TeV and PeV events. By taking into account the error ranges of $E'_{\rm LV}$ and $t'_{\rm in}$, we can draw an error band on the $\Delta t_{\rm obs}/(1+z)$ versus the $K$ plot, as shown in Fig.~\ref{fig1}. The opposite signs of LV correction also indicate that neutrinos and antineutrinos have different propagation propertie; i.e., one is superluminal and the other is subluminal. This can be explained by the Lorentz violation due to the $CPT$ odd feature of the linear Lorentz violation~\cite{Huang:2018}, and leads further to the $CPT$ violation between neutrinos and antineutrinos~\cite{Zhang:2018otj}.

\begin{table}[t]
	\small
	\centering
	\caption{\small\sf Parameters of northern hemisphere track events. The 19 northern hemisphere track events with associated GRBs are suggested by the IceCube Collaboration~\cite{Aartsen:2014aqy,Aartsen:2017wea}. The first one marked by $^\star$ is based on the first four-year search~\cite{Aartsen:2014aqy} and another 18 are from the search extended to three additional years from May 2012 to May 2015~\cite{Aartsen:2017wea}. The mark $^\dagger$ represents 12 events falling within the error band in Fig.~\ref{fig2}~(b).}
	\label{tab:2}
	\begin{ruledtabular}
		\centering
		\begin{tabular}{lllll}
			
			\noalign{\vspace{0.5ex}}
			GRB & $\Delta t_{\rm obs}~(\rm s)$ & $E$~(TeV) & $K$~(s$\cdot$TeV) & $\frac{\Delta t_{\rm obs}}{1+z}~(\rm s)$ \\
			\hline
			
			\noalign{\vspace{0.5ex}}
			100718A&	15$^\star$&	10&	3.36483&	4.7619\\
			
			\noalign{\vspace{0.5ex}}
			120612B&	47.71$^\dagger$&	0.54&	0.1817&	15.146\\
			
			\noalign{\vspace{0.5ex}}
			120911A&	120.94$^\dagger$&	0.98& 0.32975&	38.3937\\
			
			\noalign{\vspace{0.5ex}}
			130116A&	69.25&	2.1&	0.70661&	21.9841\\
			
			\noalign{\vspace{0.5ex}}
			130318A&	29.83$^\dagger$&	0.46&	0.15478	&9.46984\\
			
			\noalign{\vspace{0.5ex}}
			&44.58$^\dagger$&	0.32&	0.10768	&14.1524\\
			
			\noalign{\vspace{0.5ex}}
			130925B	&108.8$^\dagger$	&0.7&	0.23554	&34.5397\\
			
			\noalign{\vspace{0.5ex}}
			131029B	&50.49$^\dagger$&	0.68&	0.22881&	16.0286\\
			
			\noalign{\vspace{0.5ex}}
			131202B	&85.18&	1.7	&0.57202&	27.0413\\
			
			\noalign{\vspace{0.5ex}}
			140404B&	$-$38.49$^\dagger$&	1.1&	0.37013	&$-$12.219\\
			
			\noalign{\vspace{0.5ex}}
			140521B&	98.37$^\dagger$&	0.79	&0.26582&	31.2286\\
			
			\noalign{\vspace{0.5ex}}
			140603A	&41.35&	1.5&	0.50472&	13.127\\
			
			\noalign{\vspace{0.5ex}}
			&	$-$33.78&	2.1	&0.70661&	$-$10.7238\\
			
			\noalign{\vspace{0.5ex}}
			141029B&	$-$10.33$^\dagger$&	0.7&0.23554	&$-$3.27937\\
			
			\noalign{\vspace{0.5ex}}
			&$-$80.99$^\dagger$&	0.45&	0.15142&	$-$25.7111\\
			
			\noalign{\vspace{0.5ex}}
			150428B&	71.35&	3.2&	1.07675	&22.6508\\
			
			\noalign{\vspace{0.5ex}}
			150428D &	$-$43.69$^\dagger$&	0.54&	0.1817&	$-$13.8698\\
			
			\noalign{\vspace{0.5ex}}
			150507A&	58.24&	1.8&	0.60567&	18.4889\\
			
			\noalign{\vspace{0.5ex}}
			&	$-$74.44$^\dagger$&	0.69&	0.23217&	$-$23.6317\\			
			
		\end{tabular}
	\end{ruledtabular}
\end{table}

\begin{figure*}[t]
	\begin{center}
		\subfigure{\epsfig{figure=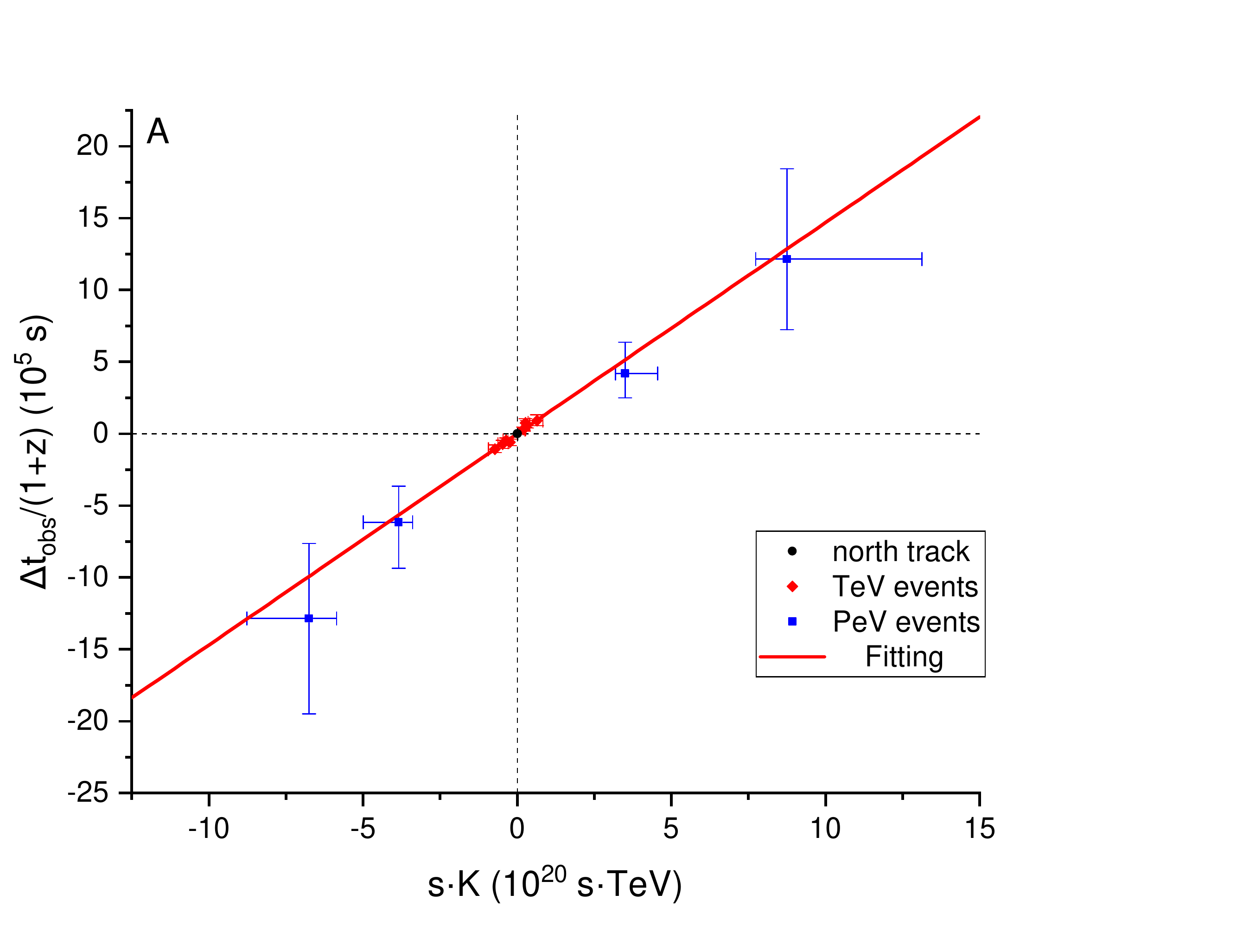, angle=0, width=8.6cm}}
		\subfigure{\epsfig{figure=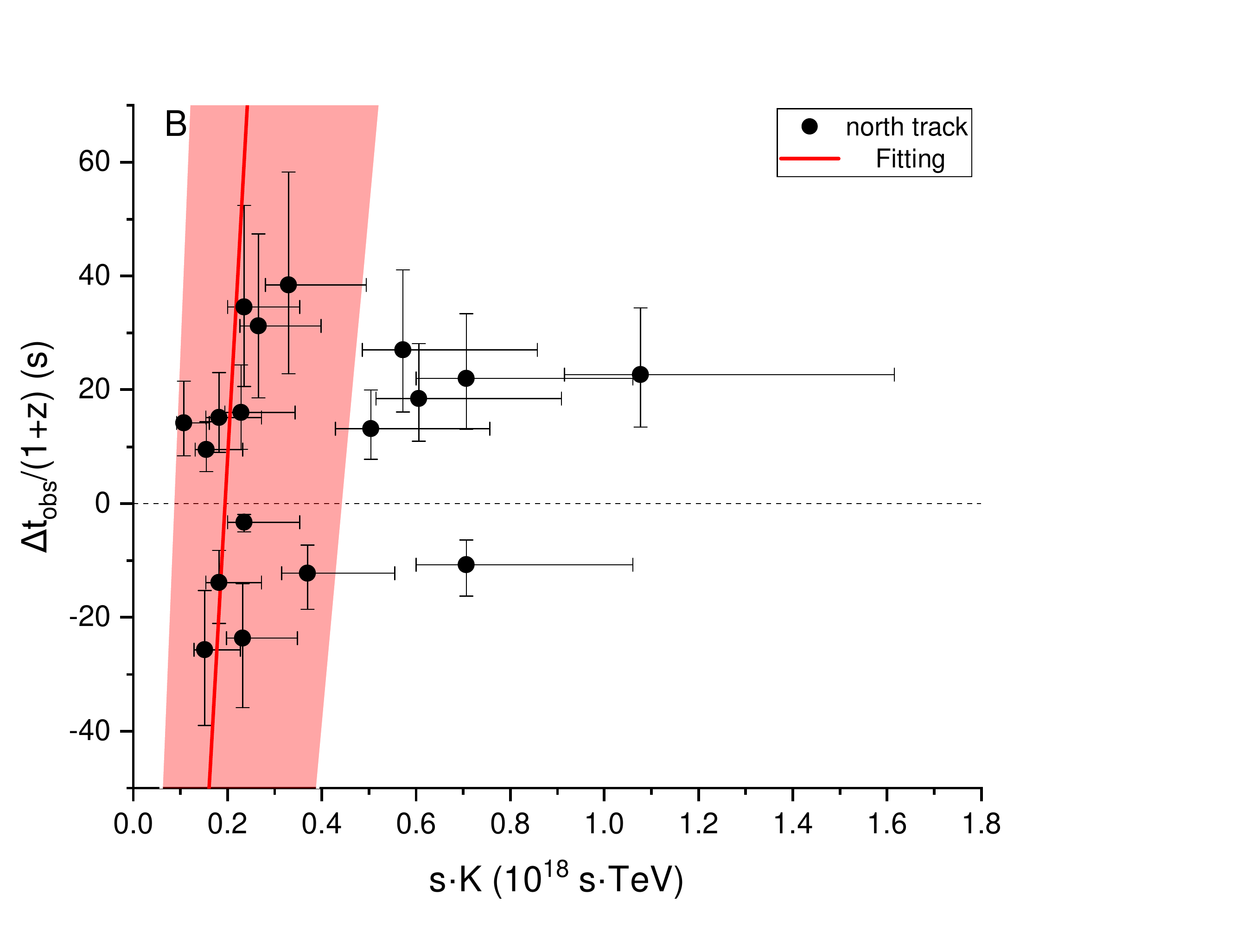, angle=0, width=8.6cm}}
	\end{center}
	\caption{Linear fitting for 18 northern hemisphere track events and high-energy events. (a) Points~(black) are experimentally measured data of northern hemisphere track events, diamonds~(red) are TeV event data, and squares~(blue) are PeV event data. Error bars are estimated according to the Methods of Ref.~\cite{Huang:2018}. The line~(red) is the fitting result. There is a well linear correlation between $\Delta t_{\rm obs}/(1+z)$ and $s\cdot K$.
		(b) The zooming in plot focuses on near-TeV events. The red colored region shows the error range of fitting. Twelve of 18 northern hemisphere track events fall within the error band of the fitting.}
	\label{fig2}
\end{figure*}

\section{Consistency in Northern Hemisphere track events}
The IceCube Collaboration~\cite{Aartsen:2014aqy,Aartsen:2016qcr,Aartsen:2017wea} also suggested associations between GRBs and lower-energy neutrino events with closer temporal coincidence. Among these events, 5  shower events are from the three-year search for all flavors of neutrinos~\cite{Aartsen:2016qcr}, one northern hemisphere track event is from the first four-year search~\cite{Aartsen:2014aqy}, another 18 are from the search extended to three additional years from May 2012 to May 2015~\cite{Aartsen:2017wea}, and 13 southern hemisphere track events are from the five-year search from May 2010 to May 2015~\cite{Aartsen:2017wea}. The southern hemisphere track events are difficult to disentangle from the atmospheric muon and $\nu_\mu+\bar{\nu}_\mu$ produced by cosmic-rays, but the atmospheric muon background from the northern hemisphere can be reduced by the Earth. Therefore in most cases, southern hemisphere events are excluded and only northern hemisphere events are analyzed~\cite{Aartsen:2017wea}.

According to Eq.~(\ref{LV factor}), we can get the LV factor $K$ of the 37 lower-energy GRB neutrino events. As recommended by the IceCube Collaboration, if the redshift value is unmeasured, we use 0.5 for short bursts and 2.15 for long bursts. We draw the $\Delta t_{\rm obs}/(1+z)$ versus the $K$ plot for all-sky shower events, northern hemisphere track events, and southern hemisphere track events in Fig.~\ref{fig1}. It is remarkable that almost all of the northern hemisphere track events fall within the error band of the energy-dependent speed variation that was revealed by the analysis on TeV and PeV GRB neutrinos~\cite{Amelino-Camelia:2016ohi,Huang:2018}. The parameters of these 19 events with associated GRBs are listed in Table~\ref{tab:2}.

\begin{figure*}[t]
	\begin{center}
		\subfigure{\epsfig{figure=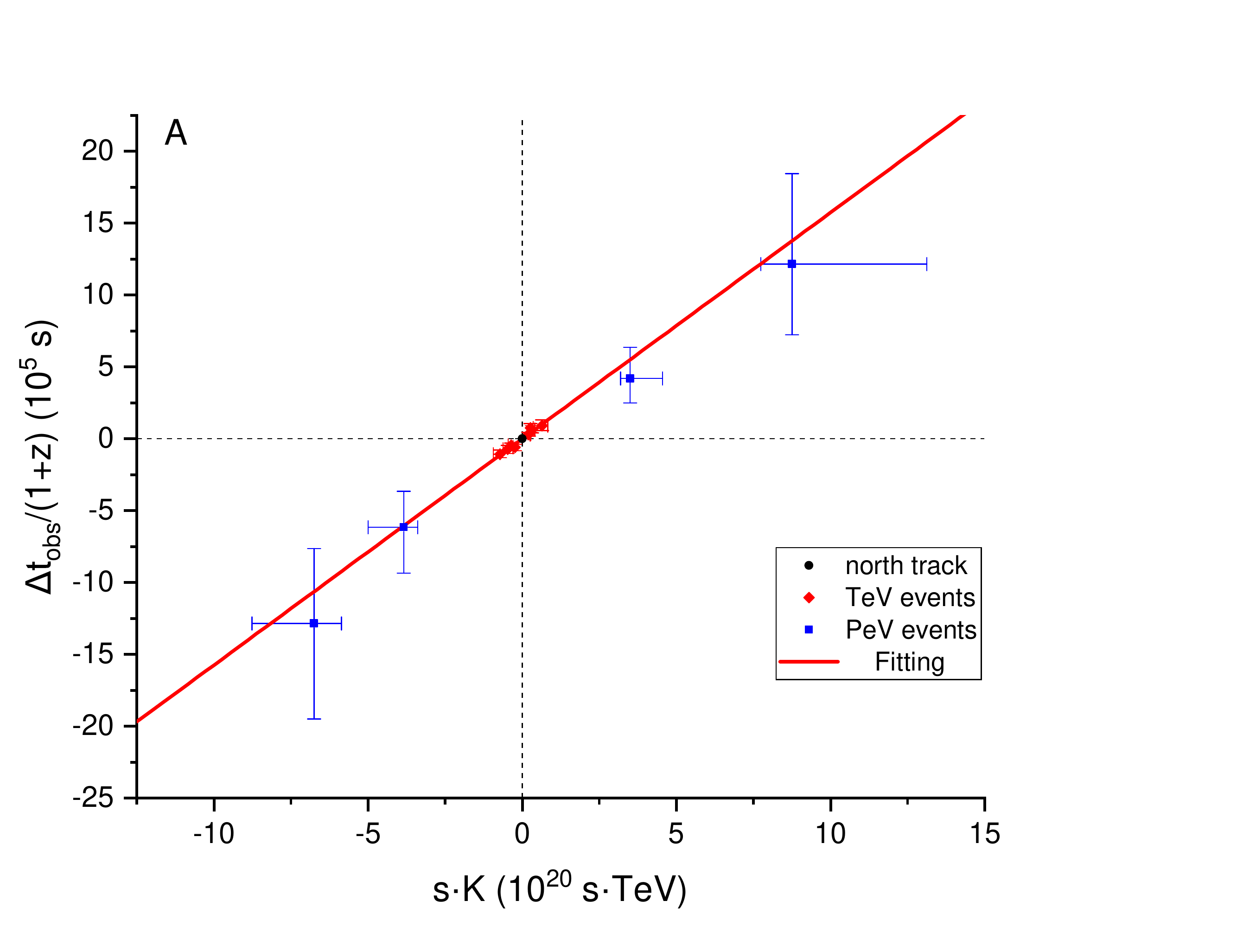, angle=0, width=8.6cm}}
		\subfigure{\epsfig{figure=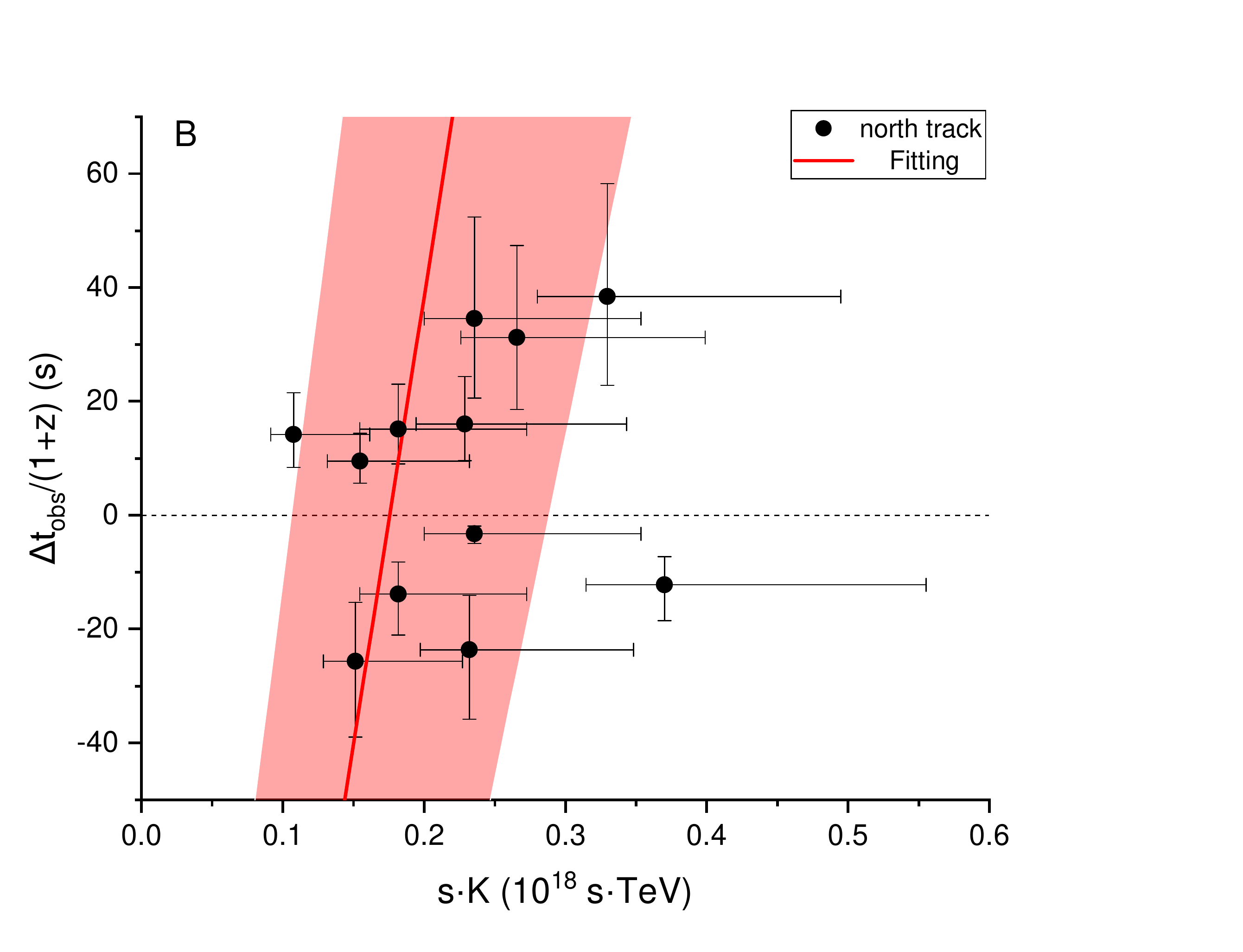, angle=0, width=8.6cm}}
	\end{center}
	\caption{Linear fitting for 12 northern hemisphere track events and high-energy events. The combined fitting indicates a Lorentz violation scale $E_{\rm LV}=(6.4\pm 1.5)\times10^{17}~{\rm GeV}$ and an intrinsic time difference $\Delta t_{\rm in}=(-2.8\pm 0.7)\times10^2~{\rm s}$. (a) Points~(black) are experimentally measured data of northern hemisphere track events, diamonds~(red) are TeV event data, and squares~(blue) are PeV event data. Error bars are estimated according to the Methods of Ref.~\cite{Huang:2018}. The line~(red) is the fitting result. There is also a well linear correlation between $\Delta t_{\rm obs}/(1+z)$ and $s\cdot K$.
		(b) The zooming in plot focuses on near-TeV events. The red colored region shows the error range of fitting.}
	\label{fig3}
\end{figure*}

To check the consistency more specifically, we do a combined fitting of 13 high-energy GRB neutrinos~\cite{Amelino-Camelia:2016ohi,Huang:2018} and 18 northern hemisphere track events from 2012 to 2015~\cite{Aartsen:2017wea}. The only event from the first four-year search and marked by $^\star$ in Table~\ref{tab:2} is excluded in our analysis, because the IceCube detector was not completed during that observation time. The error estimation is according to the Methods in Ref.~\cite{Huang:2018}. The fitting result shows a well linear correlation between the observed time difference and the LV factor, as shown in Fig.~\ref{fig2}. The fitting parameters are
\begin{eqnarray}
1/E''_{\rm LV}&=&(1.47\pm0.57)\times10^{-15}~{\rm TeV^{-1}},\label{slope2}\\
\Delta t''_{\rm in} &=&(-2.9\pm1.2)\times10^2~{\rm s},\label{intercept2}
\end{eqnarray}
and the correlation coefficient is $r=0.98$. If zooming in to the low-energy region of Fig.~\ref{fig2}~(a), as shown in Fig.~\ref{fig2}~(b), we can find that 12 of 18 northern hemisphere track events are within the error range of the combined fitting. Considering that sources of IceCube neutrino events could be more than one type, these 12 events marked by $^\dagger$ in Table~\ref{tab:2} may be really associated with GRBs. Therefore we do a new combined fitting which includes these 12 events and 13 higher-energy events as shown in Fig.~\ref{fig3}, and get the slope and the intercept
\begin{eqnarray}
1/E_{\rm LV}&=&(1.57\pm0.37)\times10^{-15}~{\rm TeV^{-1}},\label{slope3}\\
\Delta t_{\rm in} &=&(-2.8\pm0.7)\times10^2~{\rm s},\label{intercept3}
\end{eqnarray}
with the correlation coefficient $r=0.989$. Therefore the LV scale is
\begin{equation}
E_{\rm LV}=(6.4\pm1.5)\times 10^{17}~\rm{GeV},\label{LV scale3}
\end{equation}
and Eq.~(\ref{intercept3}) indicates that GRB neutrinos might be emitted about five minutes before GRB photons at the source. The new combined fitting result is well consistent with the 60~TeV to 2~PeV regularity equations~(\ref{slope1}) and (\ref{intercept1}).
Since the lower limit of these neutrino energies is several hundred GeVs, such a consistency among the gap over 4 orders of magnitude in energy scale, from a few hundred GeV to several PeV, can be a new support for the energy-dependent speed variation of ultrahigh-energy cosmic neutrinos.

\section{Neutrinos emitted before GRB photons}
The combined fitting provides us with the probability to reveal the intrinsic time difference of GRB neutrinos, since it includes neutrino events observed close to the GRB trigger time. As shown in Eq.~(\ref{intercept3}), the error range of $\Delta t_{\rm in}$ is much smaller than that in Eq.~(\ref{intercept1}), and no longer covers the zero point. It indicates clearly that GRB neutrinos might be emitted before the GRB photons. Actually, the scenario of ``early neutrino'' is supported by the evidence from astronomical observations. In the first observation of supernova neutrinos~\cite{Hirata:1987hu,Bionta:1987qt} in 1987, more than 10 neutrino events were observed several hours before the optical lights of the associated Supernova 1987A. In our discussion, $\Delta t_{\rm in}$ is only related to the intrinsic mechanism of the GRB.  It is natural to understand the negative $\Delta t_{\rm in}$ under the framework of the GRB fireball model~\cite{Eichler:1989ve,Waxman:1995vg,Waxman:1997ti,Piran:1999bk}. As a relativistically expanding plasma of electrons, photons and protons, the fireball becomes cooler and cooler during the expansion. At ultrahigh temperatures, neither neutrinos nor photons can freely escape from the intense plasma due to frequent collisions and reactions. With reduction of temperature, neutrinos and photons run out and propagate outward. Since the decoupling temperature of neutrinos is much higher than that of photons, neutrinos leak out of the dense fireball before the optical depth of photons become less than 1. Therefore, it is natural that GRB neutrinos are emitted before GRB photons. The negative $\Delta t_{\rm in}$ revealed by the combined analysis on neutrino events of different energy scales is thus supported by a rational explanation.

The regularity presented by Eqs.~(\ref{linear relation}) or (\ref{abs.}) reveals not only an energy-dependent speed variation, but also the $CPT$ reversal symmetry violation. Neutrinos and antineutrinos might have different signs of LV time correlation. For 60~TeV to 2~PeV GRB neutrinos, both ``time delay'' and ``time advance'' events exist~\cite{Huang:2018}. Although the observed time difference $\Delta t_{\rm obs}$ of northern hemisphere events in Table~\ref{tab:2} could be positive or negative, these 19 events are all time delay events. To select neutrino events, the IceCube Collaboration introduced a time probability distribution function~\cite{Aartsen:2014aqy,Aartsen:2016qcr,Aartsen:2017wea} which only includes neutrino events satisfying $-100~{\rm s}<\Delta t_{\rm obs}<300~{\rm s}$.
However, if considering a $\Delta t_{\rm in}$ of $-$280~s at the source and assuming $z=2.15$, the arrival time of the time advance events on the Earth is at least 882~s earlier than the trigger time of the associated GRB, i.e.,
\begin{equation}
\Delta t_{\rm obs}<\Delta t_{\rm in}(1+z)=-882~\rm s.
\end{equation}
Therefore, we suggest analyzing neutrino events detected a few minutes before GRBs as a method to test whether the $CPT$ violation of ultrahigh-energy neutrino does exist or not.

\section{Discussion}

In a recent study~\cite{Aartsen:2017ibm}, the IceCube
Collaboration reported some limits on the Lorentz invariance
violation by searching the anomalous neutrino oscillations of
high-energy atmospheric neutrinos under a Lorentz violation
framework of neutrino oscillations. According to the IceCube
analysis, the off-diagonal term of dimension-five LV coefficients
$\aa_{\mu\tau}^{(5)}$ was restricted to $\sim 10^{-32}\ {\rm
	GeV}^{-1}$, which seems to go beyond our result in Eqs.~(\ref{slope3}) and (\ref{LV scale3}). However, the Lorentz violation scale $E_{\rm LV}$ or its reciprocal $1/E_{\rm LV}$ in
our work represents actually the diagonal terms
$\aa_{\mu\mu}^{(5)}$ (or other flavor corresponding parameters) in
an oscillation-independent framework~\cite{Zhang:2018otj}. The
off-diagonal coefficients mainly represent the flavor transition
characters, and can thus be constrained by the measurements of LV
neutrino oscillations. But the diagonal terms have no effect on
the neutrino flavor transition, as shown by the oscillation
formula in the Methods of Ref.~\cite{Aartsen:2017ibm}. The IceCube
analysis shows that the LV is dominated by the large diagonal
component, which just corresponds to the oscillation-independent
parameters in our work. Therefore, our results are not in conflict
with the recent limits on the oscillation-dependent
parameters~\cite{Aartsen:2017ibm}.

Our analyses include both shower and track events of IceCube
neutrinos. The energy uncertainties of these two kinds of events
are different. For shower events, the energy is estimated with a
good precision. Only a part of the energy in neutral current
processes might be missed. For track events, the energy collected
by the IceCube detector may be only a part of the muon energy,
since the produced vertex of muon tracks may be located outside
the instrumental volume. The muon energy is only a part of the
total energy of the neutrino too. Therefore, the energies of track
events are reported as a lower bound. In our analyses, the
positive error of the energy is set as 30\% for shower events and 50\%
for track events, the same as the approaches in
Ref.~\cite{Huang:2018}. Actually, the true energy of neutrinos may
be even much higher than the deposited energy of track events. If we
extend the error range to a much larger value, the time window
needs to be extended too, and then more GRBs with longer observed time
differences should be taken into account. In consideration of the
narrow time window adopted by the IceCube Collaboration when
finding the correlated GRBs of the near-TeV events, we choose the
50\% positive energy error for track events as a reasonable
assumption.

The statistical significance of the near-TeV neutrino events has been analyzed in detail by the IceCube Collaboration, in the form of the
$S/B$ value~\cite{Aartsen:2016qcr,Aartsen:2017wea}. Here we use
the list of GRBs with the highest significance and choose the
possible correlations between IceCube neutrinos and GRBs. As
discussed above, because of the large amount of atmospheric $\mu$
and $\nu_\mu$ particles, these near-TeV events are still
consistent with the atmospheric neutrino backgrounds. Hence the
12 northern hemisphere track events selected by us should be
regarded as referential evidence to some extent. If these
near-TeV neutrinos are  generated from cosmic sources and really associated with
GRBs, they would be excellent supports for the energy-dependent
speed variation regularity. Even if some of these near-TeV
neutrinos are just atmospheric background, such a regularity still
cannot be ignored, since the higher-energy neutrinos at the scale
of a few tens of TeV to PeV are certainly outstanding against the
backgrounds~\cite{Amelino-Camelia:2016ohi,Huang:2018}.

\section{Summary}
In summary, by analyzing near-TeV IceCube neutrino events~\cite{Aartsen:2014aqy,Aartsen:2016qcr,Aartsen:2017wea} that are likely associated with GRBs, we find that another 12 northern hemisphere track events are consistent with the energy-dependent speed variation regularity found from 60~TeV to 2~PeV GRB neutrinos~\cite{Amelino-Camelia:2016ohi,Huang:2018}. Such a consistency over 4 orders of magnitude in energy provides a strong support of the revealed regularity.
We also suggest that GRB neutrinos are emitted several minutes before GRB photons at the source.
These findings are of fundamental importance concerning the dynamics of GRBs
and properties of neutrinos and antineutrinos.
Because of the continuous running of the IceCube Neutrino Observatory and many other detectors, these findings
are expected to be testable in the foreseeable future.

\section*{Acknowledgements} This work is supported by the National Natural Science Foundation of China (Grant No.~11475006) and the Huabao Student Research Collaborative Innovation Fund of Peking University.

\end{document}